\documentstyle[12pt]{article}
\newcommand{\be}{\begin{equation}}
\newcommand{\ee}{\end{equation}}
\newcommand{\ba}{\begin{eqnarray}}
\newcommand{\ea}{\end{eqnarray}}

\newcommand{\LL}{{\cal{L}}}

\date{}

\newcommand{\grgl}{\:\hbox to -0.2pt{\lower2.5pt\hbox{$\sim$}\hss}
           {\raise3pt\hbox{$>$}}\:}
\newcommand{\klgl}{\:\hbox to -0.2pt{\lower2.5pt\hbox{$\sim$}\hss}
           {\raise3pt\hbox{$<$}}\:}

\begin{document}
\begin{titlepage}
\begin{flushright}
HD-THEP--97--25
\end{flushright}
\bigskip
\begin{center}
{\bf\LARGE BFT Hamiltonian Embedding of}\\
\vspace{0.5cm}
{\bf \LARGE Non-Abelian Self-Dual Model}\\
\vspace{1cm}
Yong-Wan Kim\footnote{e-mail:~kim@thphys.uni-heidelberg.de

Department of Physics and Basic Science Research Institute,
Sogang University, C.P.O.Box 1142,

Seoul 100-611, Korea (ywkim@physics.sogang.ac.kr)} 
and K.D. Rothe\footnote{e-mail:~k.rothe@thphys.uni-heidelberg.de} \\
\bigskip
Institut  f\"ur Theoretische Physik\\
Universit\"at Heidelberg\\
Philosophenweg 16, D-69120 Heidelberg\\
\vspace{1cm}
\end{center}
\begin{abstract}
\noindent Following systematically the generalized Hamiltonian approach of  
Batalin, Fradkin and Tyutin, we embed the second-class non-abelian self-dual  
model of P. K. Townsend et al into a gauge theory. The strongly involutive 
Hamiltonian and constraints are obtained as an infinite power series in the 
auxiliary fields. By formally summing the series we obtain a simple 
interpretation for the first-class Hamiltonian, constraints and observables.
\end{abstract}
\vspace{8cm}
\end{titlepage}

\section{Introduction}
The quantization of second-class Hamiltonian systems \cite{Di}
 requires the strong implementation of the second-class constraints. This may  
imply Dirac brackets, whose non-canonical structure may pose problems on  
operator level. This makes it desirable to embed  the second-class theory into  
a first-class one, where the commutator relations remain canonical and the  
constraints are imposed on the states. A systematic iterative procedure  
realizing  this goal on Hamiltonian level has been developed by Batalin,  
Fradkin and Tyutin (BFT) \cite{BF,BT}. This procedure has been applied to a  
number of abelian models \cite{BRR1,BRR2,KK,BR,BB,AD,A}, where the iterative  
process terminates after a few steps. In the non-abelian case this iterative  
process may not terminate. An example is provided by the massive Yang-Mills  
theory \cite{BBN,PP}.

In this paper we consider the second-class non-abelian self-dual model of  
ref. \cite{TPV}, whose abelian version has been extensively discussed in the   
literature \cite{BRR1,BRR2,BR,AD,DJT,DJ}, and we systematically construct in  
section 2 the first-class constraints following the BFT procedure. In section  
3 we construct the observables, and in particular the Hamiltonian, as  
functionals of the first-class fields, and establish a simple relation between  
the constraints of the first- and second-class formulation. Section 4 is  
devoted to the  interpretation of the infinite power series representing the  
first-class fields, by showing that the auxiliary field in power series  
expansion plays the role of the Lie-algebra valued fields parametrizing a  
non-abelian gauge transformation. In this way we establish in section 5
the connection between the Hamiltonian BFT embedding and the corresponding  
configuration-space embedding \cite{BSV,HT}.

\section{BFT-construction of first-class constraints}

Consider the self-dual Lagrangian \cite{DJ},
\be\label{2.1}
\LL=-{1\over 2}tr f_\mu f^\mu+{\cal L}_{CS}\ee
where ${\cal L}_{CS}$ is the Chern-Simons term
\be\label{2.2}
\LL_{CS}={1\over{4m}}\epsilon_{\mu\nu\rho}tr\left(
f^{\mu\nu} f^{\rho}-{2\over 3} f^\mu f^\nu f^\rho\right).\ee
Here $f^\mu$ are (anti-hermitian) Lie-algebra valued fields
\be\label{2.3}
f^\mu=t^a f^{\mu a}\ee
and $f^{\mu\nu}$ is the usual  chromoelectric field tensor
\be\label{2.4}
f^{\mu\nu}=\partial^\mu f^\nu-\partial^\nu f^\mu +[f^\mu,f^\nu].\ee
Our conventions are
\ba\label{2.5}
[t^a,t^b]&=& c^{abc} t^c,\nonumber\\
tr(t^a t^b)&=&-\delta^{ab}.\ea
The momenta canonically conjugate to $f^{0a}$ and $f^{ia}$ are respectively  
given by $\pi^a_0=0$ and $\pi^a_i=-{1\over{2m}}\epsilon_{ij} f^{ja}$. We have  
thus two sets of primary constraints $T^a_0=0,\ T^a_i=0$, with
\ba\label{2.6}
T^a_0&=&\pi^a_0,\nonumber\\
T^a_i&=&\pi^a_i+{1\over {2m}}\epsilon_{ij} f^{ja}.\ea
The canonical Hamiltonian density associated with the Lagrangian (\ref{2.1})
is given by
\be\label{2.7}
{\cal H}_c=-{1\over 2} f^a_\mu f^{\mu a}+{1\over{2m}} f^{0a} \epsilon_{ij}  
f^{ija}.\ee
Persistency in time of these constraints leads to one further (secondary)  
constraint $T^a_3=0$, with
\be\label{2.8}
T^a_3=f^{0a}-{1\over{2m}}\epsilon_{ij} f^{ija}.\ee
The constraints (\ref{2.6}) and (\ref{2.8}) define a second-class system. In  
particular we have the Poisson-brackets
\be\label{2.9}
\left\{ T^a_i, T^b_j\right\}={1\over m} \epsilon_{ij}\delta^{ab}
\delta^2(x-y).\ee
In order to simplify the calculations, as well as for reasons that  
will become  apparent in section 5, we shall implement the constraints  
$T^a_i=0$ strongly by introducing Dirac brackets $\{\ ,\ \}_D$ defined in the  
subspace of these constraints. Following the construction of Dirac \cite{Di}, 
we have in that case, $\{ T^a_i(x),T_j^b(y)\}_D=0$, while for the remaining 
constraints one finds
\be\label{2.10}
\{\Omega^a_i(x),\Omega^b_j(y)\}_D=\Delta^{ab}_{ij}(x,y)\ee
with
\be\label{2.11}
\Delta^{ab}_{ij}(x,y)=\left(\begin{array}{cc}
0&-\delta^{ab}\\
\delta^{ab}& c^{abc}\left(-{1\over{2m}}\epsilon_{kl} f^{klc}\right)
\end{array}\right)\delta^2(x-y),\ee
where we have set $T^a_0=\Omega^a_1$ and $T^a_3=\Omega^a_2$, in order to  
streamline the notation.

We now reduce the second-class system defined by the ``commutation  
relations'' (\ref{2.10}) to a first-class system at the expense of introducing  
additional degrees of freedom. Following refs. \cite{BF,BT}, we introduce  
auxiliary fields $\Phi^{1a}$ and $\Phi^{2b}$ corresponding to $\Omega^a_1$ and  
$\Omega^a_2$, with the Poisson bracket
\be\label{2.12}
\left\{\Phi^{ia}(x),\Phi^{jb}(y)\right\}_D=\omega^{ij}_{ab}\delta^2(x-y),
\ee
where we are free \cite{BF} to make the choice
\be\label{2.13}
\omega^{ij}_{ab}=\epsilon^{ij}\delta_{ab}.\ee
The first-class constraints $\tilde\Omega^a_i$ are now constructed as a power  
series in the auxiliary fields,
\be\label{2.14}
\tilde\Omega^a_i=\Omega^a_i+\sum^\infty_{n=1}\Omega^{(n)a}_i\ee
where $\Omega^{(n)a}_i(n=1,...,\infty)$ are homogeneous polynomials in the  
auxiliary fields $\{\Phi^{jb}\}$ of degree $n$, to be determined by the  
requirement that the constraints $\tilde\Omega^a_i$ be strongly involutive:
\be\label{2.15}
\left\{\tilde\Omega^a_i(x),\tilde\Omega^b_j(y)\right\}_D=0.\ee
Making the Ansatz
\be\label{2.16}
\Omega^{(1)a}_i(x)=\int d^2y X^{ab}_{ij}(x,y)\Phi^{jb}(y)\ee
and substituting (\ref{2.14}) into (\ref{2.15}) leads to the condition
\be\label{2.17}
\int d^2 z d^2 z'X^{ac}_{ik}(x,z)\omega^{kl}_{cd}
(z,z')X^{bd}_{j\ell}(z',y)=-\Delta^{ab}_{ij}(x,y).\ee
For the choice (\ref{2.13}) for $\omega^{ij}_{ab}$, equation (\ref{2.17}) has  
(up to a natural arbitrariness) the solution
\be\label{2.18}
X^{ab}_{ij}(x,y)=\left(\begin{array}{cc}
\delta^{ab}&0\\
{1\over {4m}}c^{abc}\epsilon_{kl} f^{ijc}&\delta^{ab}\end{array}\right)
\delta^2(x-y).\ee
Substituting (\ref{2.18}) into (\ref{2.16}) as well as (\ref{2.14}), and  
iterating this procedure one finds the strongly involutive first-class  
constraints to be given by
\ba
\tilde\Omega^a_1&=&\pi^a_0+\Phi^{1a}\label{2.19}\\
\tilde\Omega^a_2&=& f^{0a}-{1\over {2m}}\epsilon_{ij} f^{ija} +\Phi^{2a}\nonumber\\
&&-\sum^\infty_{n=1}{{(-1)^n}\over{(n+1)!}}
\left[\left(\tilde\Phi^1\right)^n\right]^{ab}
\left({1\over{2m}}\epsilon_{ij}f^{ijb}\right),\label{2.20}\ea
where
\be\label{2.21}
\left(\tilde\Phi^1\right)^{ab}=c^{acb}\Phi^{1c}.\ee
It turns out convenient to define the field
\be\label{2.22}
V(\theta)=1+\sum^\infty_{n=1}{{(-1)^n}\over{(n+1)!}}
\tilde\theta^n\ee
where $\tilde\theta$ is a Lie-algebra valued field in the adjoint representation,  
$\tilde\theta=\theta^aT^a$, with $T^c_{ab}=c^{acb}$. In terms of $V(\theta)$ the  
constraint $\tilde\Omega^a_2$ reads
\be\label{2.23}
\tilde\Omega^a_2=f^{0a}+\Phi^{2a}-V^{ab}(\Phi^1)
\left({1\over{2m}}\epsilon_{ij} f^{ijb}\right).\ee
This completes the construction of the first-class constraints.

\section{Construction of first-class fields}
\setcounter{equation}{0}

The construction of the first-class Hamiltonian $\tilde H$ can be done
along similar lines as in the case of the constraints, by representing it as  
a power series in the auxiliary fields and requiring
$\{\tilde\Omega^a_i,\tilde H\}_D=0$ subject to the condition $\tilde H[f,\Phi=0]=H_c$.
We shall  follow here  a somewhat different
path \cite{KK} by noting that any functional of first-class fields $\tilde  
f^\mu$ will also be first-class.
This leads us to the identification $\tilde H=H_c[\tilde f]$. The  
``physical'' fields $\tilde f^\mu$ are obtained as a power series in the  
auxiliary fields $\Phi^{ia}$ by requiring them to be strongly involutive:  
$\{\tilde\Omega^a_i,\tilde f^\mu\}_D=0$. The iterative solution of these  
equations involves the use of (\ref{2.13}) and (\ref{2.18}) and leads to an  
infinite series which  can be compactly written in terms of $\tilde\Phi^1$  
defined in
(\ref{2.21}) as
\ba\label{3.1}
\tilde f^{0a}&=& f^{0a}+\Phi^{2a}+\left(U^{ab}(\Phi^1)-V^{ab}(\Phi^1)\right)
\left({1\over{2m}}\epsilon_{ij} f^{ijb}\right),\nonumber\\
\tilde f^{ia}&=& U^{ab}(\Phi^1) f^{ib}+V^{ab}(\Phi^1)\partial^i\Phi^{1b},\ea
where $V(\theta)$ has been defined in (\ref{2.22}) and $U(\theta)$ is given by
\be\label{3.2}
U(\theta)=1+\sum^\infty_{n=1}{{(-1)^n}\over{n!}}\tilde\theta^n.\ee
For $\tilde\pi^a_0$ we have correspondingly
\be\label{3.3}
\tilde\pi^a_0=\pi^a_0+\Phi^{1a}.\ee
We now observe that the first-class constraints (\ref{2.19}, \ref{2.20}) can  
be written in terms of the physical fields as
\ba\label{3.4}
\tilde\Omega^a_1&=&\tilde\pi^a_0\nonumber\\
\tilde\Omega^a_2&=&\tilde f^{0a}-{1\over{2m}}\tilde f^{0a}\epsilon_{ij}\tilde  
f^{ija}.\ea
Comparing with the second-class constraints $T^a_0$ and $T^a_3$ in  eqs.  
(\ref{2.6}) and (\ref{2.8}), we see that the first-class constraints  
(\ref{3.4}) are just the second-class constraints written in terms of the  
physical variables. Correspondingly, we take the first-class Hamiltonian  
density $\tilde{\cal H}$ to be given by the second-class one (\ref{2.7}),  
expressed in terms of the physical fields:
\be\label{3.5}
\tilde{\cal H}=-{1\over 2}\tilde f^a_\mu \tilde f^{\mu a}+{1\over{2m}}\tilde  
f^{0a}\epsilon_{ij}\tilde f^{ija}.\ee
It is important to notice that any Hamiltonian weakly equivalent to  
(\ref{3.5}) describes the same physics since the observables of the first-class  
formulation must be first-class themselves. Hence we are free to add to  
$\tilde{\cal H}$ any terms proportional to the first-class constraints.

\section{Interpretation of the results}
\setcounter{equation}{0}

For what follows it will be convenient to rewrite the constraints  
(\ref{2.19}) and (\ref{2.20}), as well as $\tilde f^{\mu a}$ in (\ref{3.1}) in  
terms of canonically conjugate fields. To this end we observe that the  
symplectic structure (\ref{2.12}) allows for the identifications 
$\Phi^{1a}=\theta^a, \Phi^{2a}=\pi^a_\theta$, with 
$(\theta^a,\pi^a_\theta)$ canonically conjugate pairs. 
In this notation the constraints $\tilde\Omega_i^a\approx0$ and the
fields $\tilde f^{\mu a}$ take, respectively, the form
\ba\label{4.2}
&&\tilde\Omega^a_1=\pi^a_0+\theta^a,\nonumber\\
&&\tilde\Omega_2^a=f^{0a}+\pi^a_\theta-V^{ab}(\theta)(
\frac{1}{2m}\epsilon_{ij}f^{ijb}),\ea
and
\ba\label{4.3}
&&\tilde f^{0a}=f^{0a}+\pi^a_\theta+(U^{ab}(\theta)-V^{ab}(\theta))
(\frac{1}{2m}\epsilon_{ij}f^{ijb}),\nonumber\\
&&\tilde f^{ia}=U^{ab}(\theta)f^{ib}+V^{ab}(\theta)\partial^i
\theta^b.\ea
For the first-class field strength tensor $\tilde f^{ija}$ one
has correspondingly
\be\label{4.4}
\tilde f^{ija}=U^{ab}(\theta)f^{ijb}.\ee
The field $\tilde f^{ia}$ has a simple interpretation. 
Defining the group valued field
\be\label{4.5}
g(\theta)=e^\theta,\quad \theta=\theta^at^a\ee
we have for a Lie algebra valued field $A=A^at^a$,
\ba\label{4.6}
-tr(t^ag^{-1}(\theta)Ag(\theta))=U^{ab}(\theta)A^b\nonumber\\
-tr(t^ag^{-1}(\theta)\partial_\mu g(\theta))
=V^{ab}(\theta)\partial_\mu\theta^b.\ea
The l.h.s. of these equations resumes in compact form the
infinite series on the r.h.s..
We thus see that the expression for $\tilde f^i=\tilde f^{ia}t^a$ in
(\ref{4.3}) can correspondingly be written in the compact form
\be\label{4.7}
\tilde f^{i}=g^{-1}f^{i}g+g^{-1}\partial^ig.\ee
The fields $\tilde f^{ia}$ are thus
identified with the gauge-transform
of the fields $f^{ia}$. They are invariant under the extended
gauge transformation
\ba\label{4.8}
&&f^{i}\to h^{-1}f^{i}h+h^{-1}\partial^ih\nonumber\\
&&g\to h^{-1}g\ea
and thus represent natural observables in the extended space.
Correspondingly, the first-class field strength tensor
$\tilde f^{ij}=\tilde f^{ija}t^a$ takes the form
\be\label{4.9}
\tilde f^{ij}=g^{-1}f^{ij}g.\ee
Since the field strength tensor transforms homogeneously
under gauge transformations, this is in agreement with our expectations
and suggests that we should have the weak equality
\be\label{4.10}
\tilde f^0\approx g^{-1}f^0g+g^{-1}\partial^0g.\ee
We now demonstrate this. To this end we observe that $\tilde f^{0a}$
in (\ref{4.3}) can be written as
\be\label{4.11}
\tilde f^{0a}=U^{ab}(\theta)(\frac{1}{2m}\epsilon_{ij}f^{ijb})
+\tilde\Omega_2^a.\ee
Hence recalling (\ref{4.6}), we have
\be\label{4.12}
\tilde f^0\approx g^{-1}(\frac{1}{2m}\epsilon_{ij}f^{ij})g.\ee
From (\ref{4.9}) we see that this is nothing but the
second-class constraint (\ref{2.8}) with $f^\mu$
replaced by $\tilde f^\mu$.

The above considerations indicate that $\tilde f^\mu$ is
weakly equal to nothing but the gauge transform of $f^\mu$, with $g$ in
the gauge group. In order to put this claim on a solid basis, we consider
the gauge-transform of the Lagrangian (\ref{2.1}) as given by
\be\label{4.13}
\hat{\cal L}=-\frac{1}{2}tr(\hat f^\mu\hat f_\mu)+\frac{1}{4m}\epsilon_{\mu\nu\rho}
tr(f^{\mu\nu}f^\rho-\frac{2}{3}f^\mu f^\nu f^\rho)
\ee
with $\hat f^\mu=g^{-1}f^\mu g+g^{-1}\partial^\mu g$, where we have
made use of the fact that the gauge transformation leaves
the Chern-Simons action invariant, up to a topological term and a
surface term:
\be\label{4.14}
\hat{\cal L}_{CS}(f,g)={\cal L}_{CS}(f)-\frac{2}{3}
tr(g^{-1}dg)^3+2\epsilon_{\mu\nu\rho}\partial^\rho(f^\mu\partial
^\nu gg^{-1}).\ee
Making use of (\ref{4.6}) we may write $\hat f^\mu$ in the
form
\be\label{4.15}
\hat f^{\mu a}=U^{ab}(\theta)f^{\mu b}+V^{ab}(\theta)
\partial^\mu\theta^b.\ee
For the momenta canonically conjugate to $f^{\mu a}$ and $\theta^a$
one finds
\ba\label{4.16}
&&\pi_0^a=0,\quad \pi^a_i=-\frac{1}{2m}\epsilon_{ij}f^{ja}\nonumber\\
&&\pi^a_\theta=[U^{bc}(\theta)f^{0c}+V^{bc}(\theta)\partial^0
\theta^c]V^{ba}(\theta).\ea
The first two relations represent primary constraints.
Note that the canonical momenta and fields are not to be confused
with those of the second-class formulation. The Hamiltonian
corresponding to (\ref{4.13}) takes the form
\ba\label{4.17}
\hat{\cal H}&=&\frac{1}{2m}f^{0a}\epsilon_{ij}f^{ija}+\frac{1}{2}
(U^{ab}(\theta)f^{ib}+V^{ab}(\theta)\partial^i\theta^b)^2\nonumber\\
&&+\frac{1}{2}(\pi^b_\theta(V^{-1}(\theta))^{ba})^2
-\pi^b_\theta(V^{-1}(\theta))^{ba}
U^{ac}f^{0c}.\ea
Persistency in time of the primary constraint
$\hat\Omega^a_1=\pi^a_0=0$ leads to the secondary constraint
\be\label{4.18}
\hat\Omega^a_2=\pi^b_\theta(V^{-1}(\theta))^{bc}U^{ca}(\theta)-\frac{1}
{2m}\epsilon_{ij}f^{ija},\ee
while a similar requirement for the constraints $\pi^a_i+\frac{1}{2m}
\epsilon_{ij}f^{ja}=0$ generates no further constraints,
but merely serves to fix the corresponding Lagrange multipliers
in the total Hamiltonian. As before we implement these constraints
strongly. With respect to the corresponding Dirac brackets, the other  
constraints $\hat\Omega^a_i\approx 0,\quad i=1,2$ are first-class,
and thus reflect the underlying gauge invariance of the Lagrangian
(\ref{4.13}). We now return to (\ref{4.15}).
From (\ref{4.15}) it immediately follows that
\be\label{4.19}
\hat f^{ia}=\tilde f^{ia}\ee
Making use of (\ref{4.16}) in order to eliminate $\partial^0\theta^a$
in favor of $\pi^a_\theta$, we further have
\be\label{4.20}
\hat f^{0a}=\pi^b_\theta(V^{-1}(\theta))^{ba}=
U^{ab}(\theta)(\frac{1}{2m}\epsilon_{ij}f^{ijb})+U^{ab}(\theta)\hat
\Omega_2^b\ee
or equivalently
\be\label{4.21}
\hat f^0=g^{-1}(\frac{1}{2m}\epsilon_{ij}f^{ij}+\hat
\Omega_2)g.\ee
Hence, comparing with (\ref{4.12}), we conclude
\be\label{4.22}
\hat f^0\approx \tilde f^0\ee
This establishes the weak equivalence of $\hat f^\mu$
and $\tilde f^\mu$. We furthermore have from (\ref{4.18})
\be\label{4.23}
V^{ab}(\theta)\hat\Omega^b_2=\pi^a_\theta-
V^{ab}(\theta)(\frac{1}{2m}\epsilon_{ij}
f^{ijb}),\ee
where we have made use of the identity
\be\label{4.24}
U^{ac}(\theta)V^{bc}(\theta)=V^{ab}(\theta).\ee
Performing the canonical transformation
\ba\label{4.25}
&&\pi^a_\theta\to\pi^a_\theta+f^{0a}\nonumber\\
&&\pi^a_0\to\pi^a_0+\theta^a\ea
the first-class constraints $\hat\Omega^a_0\approx0$
and $V^{ab}\hat\Omega^b_3\approx 0$
map into the constraints (\ref{4.2}). It remains to check the
relation between $\hat{\cal H}$ and $\tilde{\cal H}$. Making use of
(\ref{4.20}), expression (\ref{4.17}) for $\hat{\cal H}$
may be written in the form
\ba\label{4.26}
\hat{\cal H}&=&\frac{1}{2}(U^{ab}(\theta)f^{ib}+V^{ab}(\theta)
\partial^i\theta^b)
^2+\frac{1}{2}(\frac{1}{2m}\epsilon_{ij}f^{ija})^2\nonumber\\
&&+\frac{1}{2}(\hat\Omega^a_3)^2+(\frac{1}{2m}\epsilon_{ij}f^{ija}-
f^{oa})\hat\Omega^a_3.\ea
Comparison of (\ref{4.26}) with (\ref{3.5}) immediately shows
that $\hat{\cal H}\approx\tilde{\cal H}$. We thus conclude
that the BFT constructioon is equivalent to the quantization of
the gauge theory defined by the Lagrangian (\ref{4.13}).

\section{Revisiting section 4}
\setcounter{equation}{0}
The discussion of section 4 indicates that there exists a more
economical path for arriving at the results. It consists in gauging
the Lagrangian (\ref{2.1}) by making the substitution
$f^\mu\to\hat f^\mu$. The resulting Lagrangian, eq. (\ref{4.13}),
can be written in the form
\be\label{5.1}
\hat{\cal L}(f,g)={\cal L}(f)+{\cal L}_{WZ}\ee
where
\be\label{5.2}
{\cal L}_{WZ}(f,g)=-tr(f^\mu\partial_\mu gg^{-1})-\frac{1}{2}tr
(g^{-1}\partial^\mu g)^2\ee
plays the role of the Wess-Zumino-Witten (WZW) term in
the gauge-invariant formulation of two-dimensional chiral gauge
theories \cite{BSV, HT, AAR}, and ${\cal L}(f)$ is the Lagrangian
of the second-class system.
Contrary to what was done in section 4, we choose to work here with
the group valued field $g$, instead of the Lie-algebra valued
field $\theta$. We then have for the momentum $\Pi$
conjugate to $g$,
\be\label{5.3}
\Pi^T=-g^{-1}f^0-g^{-1}\partial^0gg^{-1}\ee
where ``T'' denotes ``transpose''. The canonical momenta
$\pi_\mu$ are the same as before. Hence the primary constraints
are still of the form (\ref{2.6}),
\be\label{5.4}
\hat T_0^a=\pi_0^a,\quad \hat T_i^a=\pi^a_i+
\frac{1}{2m}\epsilon_{ij}f^{ija}\ee
though the dynamics is a different one. The canonical Hamiltonian
corresponding to (\ref{5.1}) reads, on the constraint surface
$\pi^a_0=0$,
\ba\label{5.5}
H_C&=&\int d^2x\left\{
-\frac{1}{2}tr(\Pi^Tg)^2-\frac{1}{2}tr f^2_i + tr(f^ig\partial_ig^{-1})
\right.\nonumber\\
&&\left.-\frac{1}{2}tr(g^{-1}\partial^ig)^2+tr f^0(\partial^i\hat
T_i+\hat T_3)\right\}\ea
where $\hat T_3$ is given by
\be\label{5.6}
\hat T_3=-\frac{1}{2m}\epsilon_{ij}f^{ij}-g\Pi^T.\ee
The requirement $\dot\pi^a_0=0$ leads to the secondary
constraint $\hat T_3^a + \partial^i \hat T_i=0$.

The constraints $\hat T_i^a=0$ are evidently second class,
and as before we implement them strongly by working with the
corresponding Dirac brackets. On the surface defined by 
$\hat T_i=0$, the secondary constraints read $\hat T^a_3=0$.
One easily checks that the constraints $\hat T^a_0=0$ and
$\hat T^a_3=0$ are first class with respect to these Dirac brackets.

It remains to establish the relation with the results
of section 4. Multiplying (\ref{5.3}) from the right with
$g$ and using (\ref{4.6}) we have
\ba\label{5.7}
tr(t^ag\Pi^T)&=&f^{0a}+V^{ab}(-\theta)\partial^0\theta^b\nonumber\\
&=&f^{0a}+U^{ca}(\theta)V^{cb}(\theta)\partial^0\theta^b,\ea
where we have further used
\be\label{5.8}
U^{ca}(\theta)V^{cb}(\theta)=V^{ab}(-\theta).\ee
Comparing (\ref{5.7}) with (\ref{4.16}) and making further
use of
\be\label{5.9}
U^{ba}(\theta)U^{ca}(\theta)=\delta^{ab}\ee
we see that
\be\label{5.10}
V^{ac}(\theta)U^{ab}(\theta)tr(t^bg\Pi^T)=\pi^c_\theta\ee
On the other hand, from (\ref{5.6}) we deduce
\be\label{5.11}
U^{ab}(\theta)tr(t^bg\Pi^T)=-U^{ab}(\theta)
(\frac{1}{2m}\epsilon_{ij}f^{ija}
+\hat T_3^a).\ee
Let us compare this with $\hat f^0$ defined by
\be\label{5.12}
\hat f^0=g^{-1}f^0g+g^{-1}\partial^0g.\ee
From (\ref{5.3}) we see that
\be\label{5.13}
\hat f^0=-\Pi^Tg.\ee
Using (\ref{5.6}) and noting that
$\hat T_0=\hat\Omega_1, \hat T_3=\hat\Omega_2$, we recover
(\ref{4.21}). This establishes the equivalence of the various
procedures.

\section{Conclusion}

The main objective of this paper was to provide a nontrivial example
for the Hamiltonian embedding of a second-class theory
into a first-class one, following the systematic constructive
procedure of Batalin, Fradkin, and Tyutin (\cite{BF,BT}).
Unlike the case of the abelian models discussed in the literature,
the first-class Hamiltonian and secondary constraint generated
by this procedure are obtained as an infinite power series
in the auxiliary fields living in the extended phase space.
By explicitly summing this series we established the weak
equivalence with the corresponding quantities as obtained
by gauging the second-class Lagrangian defining our model, with
the auxiliary fields playing the role of the corresponding
gauge degrees of freedom. We thereby showed
that on the space of gauge-invariant functionals the Lagrangian
approach of refs. \cite{BSV,HT} for embedding second-class
theories into a gauge theory is equivalent to the Hamiltonian
BFT approach. We further showed that the most economical
way of obtaining the desired results would consist in working
with the group rather than Lie-algebra valued fields of the
gauged Lagrangian. One readily checks that the same conclusion
can be drawn for the model of ref. \cite{BBN}.

\section*{Acknowledgement}
We would like to thank R. Banerjee and Y. J. Park for
helpful comments. One of the authors (Y. W. Kim) would
like to thank the Institut f\"ur Theoretische Physik for the
kind hospitality and the Korea Research Foundation for  
(1996) overseas fellowship, 
which made this collaboration possible.

\end{document}